# Gravitational Wave Astrophysics: Opening the New Frontier

Joan Centrella[1]

*Astrophysics Science Division, NASA's Goddard Space Flight Center, Greenbelt, MD 20771, USA*

**Abstract.** The gravitational wave window onto the universe is expected to open in ~ 5 years, when ground-based detectors make the first detections in the high-frequency regime. Gravitational waves are ripples in spacetime produced by the motions of massive objects such as black holes and neutron stars. Since the universe is nearly transparent to gravitational waves, these signals carry direct information about their sources – such as masses, spins, luminosity distances, and orbital parameters – through dense, obscured regions across cosmic time. This article explores gravitational waves as cosmic messengers, highlighting key sources, detection methods, and the astrophysical payoffs across the gravitational wave spectrum.

## INTRODUCTION

Gravitational waves (GWs) are a remarkable prediction of Einstein's General Relativity. They are generated by the dynamics of massive of objects, such neutron star (NS) binaries and black hole mergers. And although we have excellent *indirect* evidence of GW emission through its effects on the orbital evolution of the Hulse-Taylor pulsar B1913+16 and the double pulsar J0730-3039B, we do not yet have a direct detection of GWs. This situation is poised to change dramatically by mid-decade, when we expect the first observations of GWs from compact binaries will be made by ground-based GW detectors.

GWs are distinctive cosmic messengers. Gravitational waveforms encode the dynamics of their sources, and thus provide the source masses, spins, orbits, and luminosity distances directly. GWs couple very weakly to matter, so the waveforms carry this information virtually unscathed, through dense regions that obscure electromagnetic (EM) radiation, across the span of cosmic time.

GW sources can be extremely energetic. Consider the final merger of a comparable mass black hole binary as an example. The GW luminosity produced when the black holes merge is ~ $10^{23}$ $L_{SUN}$, where $L_{SUN}$ is the solar luminosity. This is far more energy than a gamma-ray burst (GRB) emits in EM radiation, and is actually greater than the energy in EM radiation emitted by all the stars in all the galaxies in the observable universe, during the time of the black hole merger.

In this article, we introduce this emerging field of GW astrophysics. We begin with some basic properties of GWs, and then discuss the full GW spectrum, highlighting key sources and detectors in each frequency band. Next, we compare the properties of EM radiation and GWs as cosmic messengers, discussing their differences and complementarities. We then give an overview of GW detection, including a discussion of basic principles and a survey of current detection efforts across the GW spectrum. The next section features key binary sources of GWs, and includes discussion of the expected rates of detection. We conclude with a discussion of multi-messenger astrophysics combining EM and GW observations, and the excitement expected as the GW window onto the universe opens.

We have endeavored to write this article at a level accessible without specialized knowledge of GWs and their detection. While this material is written as an overview, it is beyond the scope of this article to provide a comprehensive list of references. We have instead referred to key review papers where the reader can find more details and references.

## BASIC PROPERTIES OF GRAVITATIONAL WAVES

According to General Relativity, spacetime is curved due to the presence of mass. When masses move, they cause disturbances in spacetime that propagate outward from their source, carrying energy and momentum. Far from their source, these ripples are small perturbations $|h_{\mu\nu}| \ll 1$ ($\mu,\nu = 0,1,2,3$) on a flat background spacetime. Using suitable coordinates and gauge choices [1,2], Einstein's equations become the wave equation

$$\left(\nabla^2 - \frac{1}{c^2}\frac{\partial^2}{\partial t^2}\right)h_{\mu\nu} = 0, \tag{1}$$

so these ripples $h_{\mu\nu}$ are a wave phenomenon, called gravitational waves, which propagate at the speed of light $c$. With these coordinate and gauge choices, the radiative degrees of freedom appear only in the spatial components $h_{ij}$ ($i,j = 1,2,3$). Physically, there are only two independent radiative degrees of freedom, corresponding to two polarization states which are called $h_+$ and $h_x$; a general GW is a superposition of these two polarizations. When discussing general properties of GWs, we denote the waves simply as $h$. Note that $h = h(t - r/c)$ and $h \sim 1/r$, where $r$ is the distance from the source.

GWs are purely transverse, producing differential gravitational accelerations in a plane perpendicular to their direction of propagation. This property forms the basis for GW detectors, as we discuss below. Figure 1 shows the lines of force from plane GWs traveling along the $z$-axis.

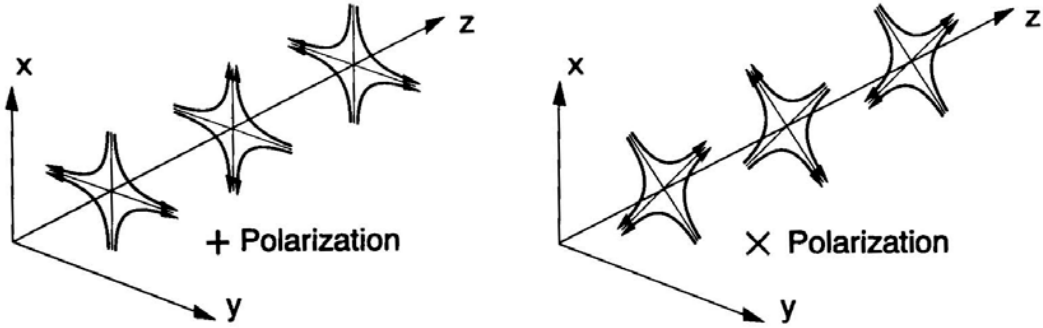

**FIGURE 1.** Lines of force for plane GWs traveling along the $z$-axis. The left panel shows $h_+$ and the right panel shows $h_\times$. These gravitational waves produce tidal accelerations in the $x$-$y$ plane. A general gravitational wave is a superposition of $h_+$ and $h_\times$. This figure is from Ref. [3] and is reprinted with permission from AAAS.

By analogy with EM waves, we can decompose GWs into multipolar contributions that reflect the nature of the source that generates them. Conservation of mass requires that there be no monopole gravitational radiation, and conservation of momentum rules out dipole gravitational radiation [1]. The leading order contribution is then quadrupolar: GWs are generated by sources with time-changing quadrupole moments, such as binaries. For a source of total mass $M$ located at distance $r$, we can write the dimensionless amplitude of the GW as

$$h \sim \frac{G}{c^4} \frac{\ddot{Q}}{r} \sim \frac{R_{Sch}}{r} \frac{v^2}{c^2}, \qquad (2)$$

where $Q$ is the (traceless) quadrupole moment of the source, the over-dots signify time derivatives, $v$ is the characteristic non-spherical velocity in the source, and $R_{Sch} = 2GM/c^2$ is the Schwarzschild radius of the source. Eq. 2 shows that the strongest sources of gravitational radiation have large masses moving at velocities near $c$.

Binaries consisting of compact objects such as black holes or NS are generally expected to be the strongest sources of GWs. While all binaries emit GWs, the amplitude of this radiation is so weak compared to other effects, such as dynamical friction from stellar encounters and gaseous dissipation, that it only becomes a dominant factor in the binary evolution when the components are highly compact and very close together. Consider a binary consisting of two black holes close to the time of final coalescence. If the black holes have individual masses $m \sim 10 M_{SUN}$ and are moving at $v \sim c/3$, then Eq. 2 gives $h \sim 10^{-20}$ if the binary is at distance $r \sim 15$ Mpc (Virgo Cluster), and $h \sim 10^{-22}$ for $r \sim 450$ Mpc (redshift $z \sim 0.1$). Such small amplitudes result from the weakness of the gravitational interaction and require precision measurement techniques for detection, as discussed below.

For binaries on quasi-circular orbits in the Newtonian quadrupole limit, the instantaneous frequency $f_{GW}$ of the emitted GWs is given by

$$f_{GW} \approx 2 f_{orb} \approx \frac{1}{\pi} \left( \frac{GM}{a^3} \right)^{1/2}, \qquad (3)$$

where $f_{orb}$ is the binary orbital frequency, $M$ is the total mass, and $a$ is the separation of the component masses. In the GW dominated regime, the binary components spiral together as the gravitational radiation carries away energy and momentum. This causes the binary orbital frequency to increase with time.

The gravitational waveforms generated by a source encode the parameters and dynamics of that source [2]. For example, at wide separations, the gravitational waveform produced by a binary is a sinusoid with (nearly) constant amplitude and (nearly) constant frequency given by Eq. 3. As the binary masses spiral closer

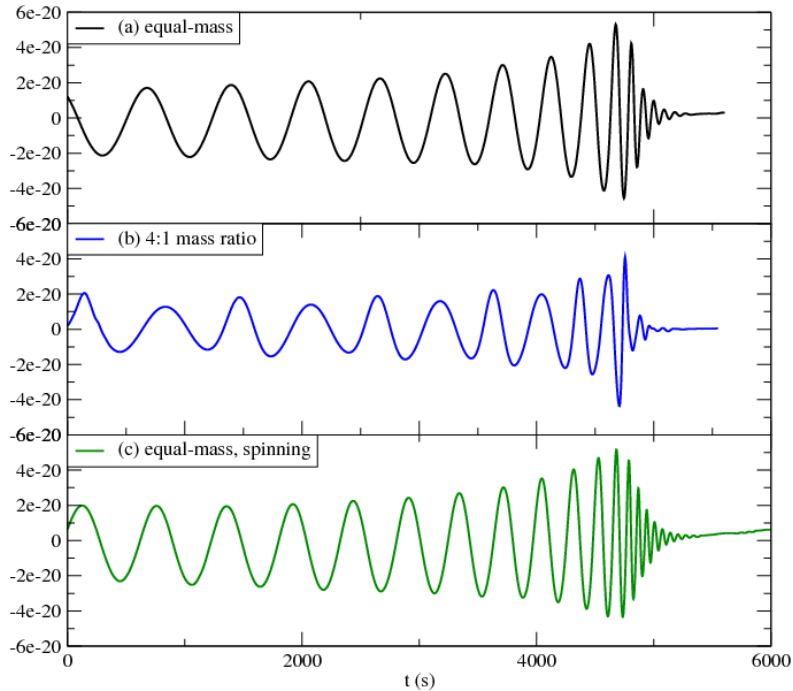

**FIGURE 2.** Gravitational waveforms calculated from binary black hole mergers. These waveforms include the last several cycles of the inspiral chirp, followed by the merger burst (where the maximum GW amplitude is reached), and then the ringdown (with increasing frequency and exponential amplitude decay). One polarization, as observed from the equatorial plane of the binary, is shown for each case. In all cases, the waveforms are scaled for a binary having total mass $M \sim 10^5 M_{SUN}$ located at redshift $z \sim 15$. Figure reprinted from Ref. [4] by permission of the publisher (Taylor & Francis Group, http://www.informaworld.com).

together, the effects of gravitational radiation reaction increase and the waveform takes on the characteristic *chirp* signature: a sinusoid increasing in both frequency and amplitude. If the binary consists of two black holes, the merger forms a single, highly perturbed black hole, which then "rings down" as GWs carry away the distortions

(much like acoustic waves carry away the distortions of a bell that has been struck) to become a quiescent rotating black hole. Figure 2 shows gravitational waveforms computed from 3-D numerical relativity simulations of merging black holes having a total system mass $M = 10^5 M_{SUN}$, as would be seen by an observer located on the equatorial plane of the binary at redshift $z \sim 15$ [4]; see also Ref [5]. Here, only a single polarization state is shown. Differences among these waveforms, reflecting the different mass ratios and spins, can be seen by eye. Detailed analysis of gravitational waveforms using parameter estimation techniques can extract system parameters, after accounting for detector noise (not shown here) [6, 7].

## THE GRAVITATIONAL WAVE SPECTRUM

Throughout the 20th century, new regions of the EM spectrum were opened up for astronomical observations, bringing a wealth of discoveries and new information. In this decade, the first *direct* observations of GWs are expected to open the GW spectrum. Figure 3 shows the full GW spectrum, ranging over $\sim$ 18 orders of magnitude in frequency. In this section, we use the properties of GWs from compact binaries and Eq. 3 to illuminate different regions of the GW spectrum.

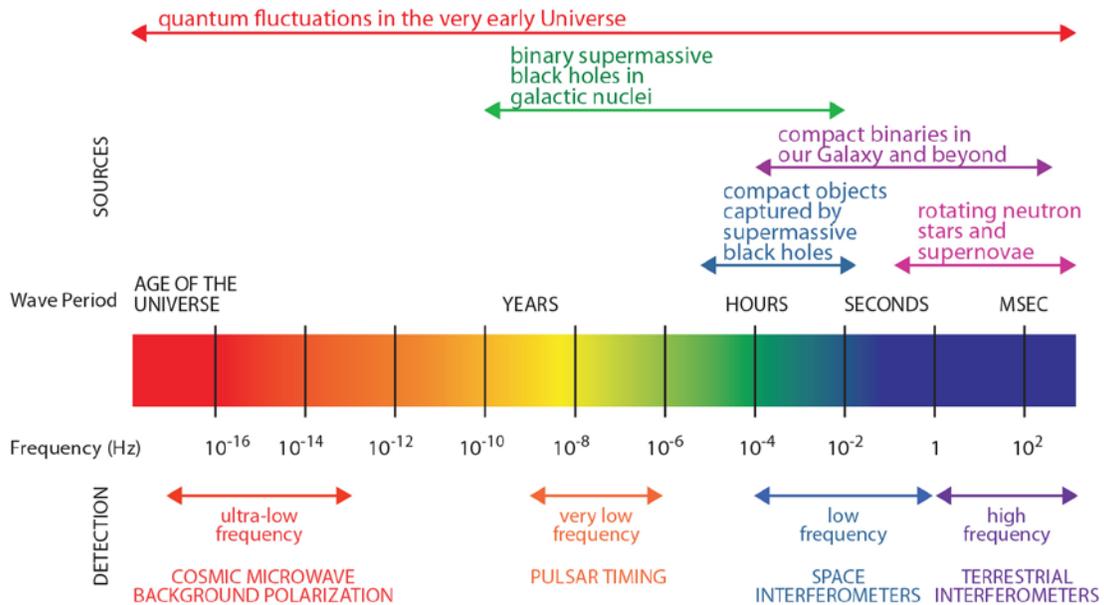

**FIGURE 3.** The GW spectrum. Figure courtesy of NASA.

The *high frequency* part of the GW spectrum spans the frequency range 1 Hz $\leq f_{GW} \leq 10^4$ Hz. The strongest sources in this band are compact binaries consisting of two NS, two stellar black holes, or a NS and a black hole. For example, a binary consisting of two black holes each having mass $\sim$ 10 $M_{SUN}$ and separated by distance $a$

$\sim 5R_{Sch}$ emits GWs with frequency $f_{GW} \sim 100$ Hz. High frequency GWs are detectable by ground-based instruments such as the US Laser Interferometer Gravitational-wave Observatory (LIGO) and the European Virgo detector, which are laser interferometers having kilometer-scale arms [8].

The *low frequency* part of the spectrum includes frequencies in the range $10^{-4}$ Hz $\leq f_{GW} \leq 10^{-1}$ Hz. This regime is the richest part of the GW spectrum, with a wealth of astrophysical sources. Mergers of massive black holes (MBHs) with masses in the range $10^4 M_{SUN} \leq M \leq 10^7 M_{SUN}$ are the strongest low frequency sources. A MBH binary with total mass $M \sim 10^6$ $M_{SUN}$ and separation $a \sim 2R_{Sch}$ emits GWs with frequency $f_{GW} \sim 2 \times 10^{-3}$ Hz. Other important low frequency sources include compact stellar binaries and extreme mass-ratio inspirals (EMRIs), in which a stellar compact object spirals into a MBH at the center of a galaxy. GWs with frequencies below $\sim 1$ Hz cannot be detected from the ground because of noise from fluctuating gravity gradients; the low frequency regime can thus only be observed from space [8]. The Laser Interferometer Space Antenna (LISA) is a proposed space-borne interferometer with arm lengths $\sim 5 \times 10^6$ km that can detect a host of low frequency GW sources.

Supermassive black holes (SMBHs), $M \geq 10^8 M_{SUN}$, at wider separations produce gravitational radiation in the *very low frequency* part of the spectrum, $10^{-9}$ Hz $\leq f_{GW} \leq 10^{-6}$ Hz. For a SMBH binary with $M \sim 10^9$ $M_{SUN}$ and separation $a \sim 500$ $R_{Sch} \sim 10^4$ AU, the frequency of emitted GWs is $f_{GW} \sim 2 \times 10^{-9}$ Hz. These systems can be observed using a pulsar timing array (PTA), which provides precise measurements of changes in the arrival times of pulsar signals due to the passage of GWs between the pulsar and the earth [9].

Table 1 summarizes the key binary GW sources and detectors for the high, low, and very low frequency bands. More details on both sources and detectors are given in later sections.

**TABLE 1.** GW frequency bands, binary sources, and detectors. In this table, WD denotes a white dwarf, BH signifies a stellar black hole with mass $M \sim 10 M_{SUN}$, MBH indicates a massive black hole with mass $M \sim (10^4 - 10^7) M_{SUN}$, and SMBH denotes a supermassive black hole with mass $M \geq 10^8 M_{SUN}$.

| Frequency Range | Key Binary Sources | Detectors |
|---|---|---|
| ***High:*** $(1 - 10^4)$ Hz | NS-NS, NS-BH, BH-BH | Ground-based interferometers: *LIGO, VIRGO, LCGT* |
| ***Low:*** $(10^{-4} - 10^{-1})$ Hz | MBH-MBH, EMRIs, Galactic Binaries (mostly WD-WD) | Space-based interferometers: *LISA, DECIGO* |
| ***Very Low:*** $(10^{-9} - 10^{-6})$ Hz | SMBH binaries (stochastic background, indiv sources) | Timing of Millisec Pulsars: *PPTA, NANOGrav, EPTA, IPTA* |

Other sources of gravitational radiation include processes in the early universe, such as quantum fluctuations, cosmic strings, phase transitions, and the dynamics of branes. These sources typically produce stochastic backgrounds (though cosmic strings can also produce bursts) spanning many frequency bands. GWs produced by cosmic strings are an interesting source for PTAs in the very low frequency regime, and may be detected in this decade. Certain models also predict that ground-based and space-based detectors may be able to observe GWs from cosmic strings. See Refs. [10, 11].

In particular, the *ultra low frequency band*, $10^{-18}$ Hz $\leq f_{GW} \leq 10^{-13}$ Hz, can be probed indirectly using measurements of the cosmic microwave background (CMB). GWs in the regime have wavelengths on the scale of the universe, $10^{-5}$ $H_0^{-1} \leq \lambda_{GW} \leq H_0^{-1}$, where $H_0^{-1}$ is the Hubble length [12]. These waves could have been generated during an inflationary epoch in the early universe, and would produce a distinctive polarization signature in the CMB. Such indirect signals are outside the scope of this article, see, e.g., Ref. [13] and references therein for more details.

## GRAVITATIONAL WAVES AS COSMIC MESSENGERS

Most of the information we have today about the universe comes in the form of EM radiation. Throughout the 20th century, the growing use of multi-wavelength astronomy greatly increased our knowledge of astronomical sources on a wide range of scales. In the early part of the 21st century we will expand this picture significantly by observing GW signals. In this section, we highlight some key differences between observing with EM and gravitational radiation.

*What information about the source does the radiation bring directly?* EM radiation is produced mainly by hot plasmas such as are found in stellar coronae, accretion disks, and intra-cluster gas. These signals directly probe the thermodynamics of these systems and their environments. In contrast, GWs are produced by the bulk motions of masses such as black holes and neutron stars. Gravitational waveforms encode these bulk motions, and bring direct information about the system dynamics such as masses, spin vectors, luminosity distances and orbits; cf. Figure 2.

*Do sources emit radiation in more than one frequency band?* Many astrophysical sources emit radiation in more than one part of the EM spectrum. Generally, the different types of EM radiation from a source come from different physical processes, often occurring at different locations within the source. For example, EM radiation from the Galactic center is emitted as radio waves (warm gas), infrared (stars and dust), and X-rays (very hot gas around the point source Sgr A*). GW emission typically has a different character, depending on the type of source. GW backgrounds from processes in the early universe can exist at varying amplitudes across much of the GW frequency spectrum. A GW background resulting from many binary sources, such as double white dwarf (WD) binaries in the Galaxy, will have a frequency range corresponding to the range of scales of all the component sources; cf. Eq. 3. For individual binaries the GWs typically are emitted in one frequency band, although the chirping (frequency increase) can cause them to evolve from, say, the low frequency into the high frequency band, as in the case of black hole binaries with $M \sim 10^3 M_{SUN}$. Of course, a very interesting possibility occurs if we can observe *both* EM and GW radiation from a source; we will return to this in a later section.

*What type of data do we obtain from observations?* The EM radiation from a source has wavelengths much smaller than the size of the emitting regions. Generally, an EM energy flux is observed, which falls off with distance to the source according to $\sim 1/r^2$. Typical products of EM observations are images and spectra of the sources. GWs generally have wavelengths either comparable to or greater than the size of their source, and do not produce images. The waveform itself is the observable, including both amplitude and phase information. And, since $h \sim 1/r$, a factor of two increase in sensitivity means that a detector can register signals from a volume nearly an order of magnitude larger. GW detectors typically produce two data streams or time series, $h_+(t)$ and $h_x(t)$ that encode the source parameters.

*What are the resolution and pointing characteristics of the detectors?* EM telescopes are generally pointed detectors with relatively high resolution. For example, Hubble Space Telescope has resolution ~ 0.05 arcsec and field-of-view (FOV) ~ 10 arcmin$^2$. The Chandra X-ray telescope has resolution ~ 1 arcsec and a $10^3$ arcmin$^2$ FOV. In contrast, GW telescopes are all-sky instruments, receiving all signals in their frequency bands that pass through the detectors. In order to localize the source on the sky some sort of triangulation is needed, either using multiple detectors and time-of-flight delays, as in the case of ground-based interferometers, or detector motion, for space-borne interferometers. Typical values of sky localization for GW sources extend from ~ 10 deg$^2$ for a network of ground-based detectors to ≤ 1 deg$^2$ for a space-borne detector such as LISA.

*How does the radiation interact with matter?* EM waves interact much more strongly with matter than do GWs; this has important consequences for both the types of signals that we can receive, and the difficulty of detecting them. For example, EM waves can be easily scattered or absorbed. These processes modify and sometimes obscure the EM signal, depending on the wavelength of the radiation and the type of intervening material. However, EM signals which do arrive at our detectors are easily recorded. In contrast, GWs are very weakly coupled to matter. Since the universe is highly transparent to GWs, we can observe these signals from sources that are at great distances and may be obscured from EM detection by gas along the line of sight. However, this small coupling means that the interaction of a GW with a detector is also very weak. From Eq. 2, a GW typically has an amplitude $h \leq 10^{-21}$ and thus will perturb any intervening matter – including a detector – by an extremely small amount. As a result, GW observations require high precision measurements, with large-scale experimental efforts essential to achieving even the first detections.

## DETECTING GRAVITATIONAL WAVES

All GW detectors today operate by monitoring changes in the light travel time between point masses caused by a passing GW. In this section we give an overview of the key ideas in GW detection, and a brief survey of current and future detectors.

### Basic Concepts in Gravitational Wave Detection

GWs traveling through spacetime produce very small deviations in the positions of isolated test masses. Since the waves cause a strain in the detector, changes in the positions of the test particles indicate the passage of a GW [8].

Ground-based GW detection uses laser interferometers to monitor the effects of a GW on the displacement of test masses. Figure 4 shows a schematic drawing of the LIGO detector. At the top of the figure, the lines of force for a GW $h_+$ traveling along the $z$-axis are shown. In the detector, the mirrors are freely suspended by pendulums and act as test masses; much effort goes into isolating them from disturbances such as seismic noise in the earth's crust and gas molecules in the surrounding air. Light from the laser is split into two beams at the beamsplitter, and sent into the two interferometer arms, bouncing back and forth between the mirrors to build up a strong signal. The two beams are then recombined at the photo detector.

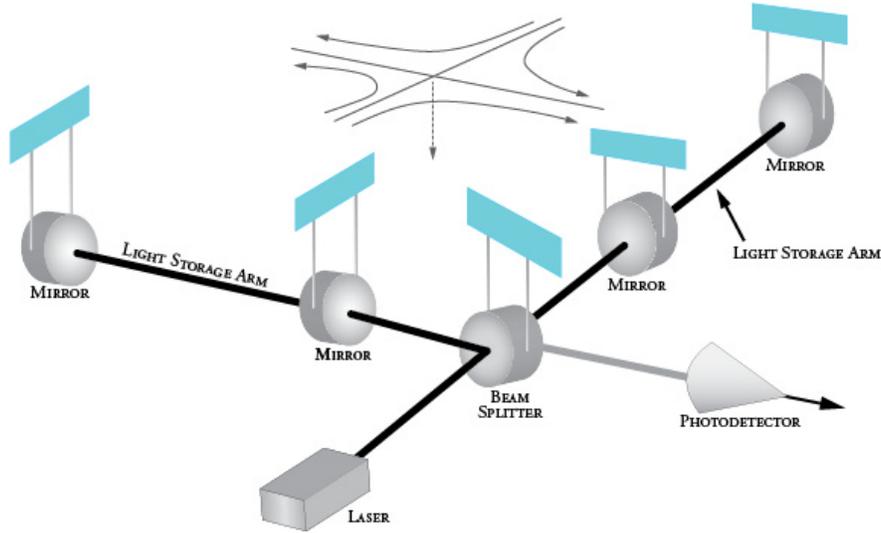

**FIGURE 4.** Schematic drawing of a ground-based laser interferometer for GW detection. The beam splitter sends light down the two arms; the beams are then interfered at the photodetector. A passing GW changes the arm lengths, causing phase changes in the laser beams. These phase changes appear in the interference patterns at the photo detector, giving the GW signal. Figure courtesy of the LIGO Laboratory, used with permission.

In Figure 4, both arms have length $L$ and are aligned with the $x$ and $y$ axes. A GW $h_+$ (with wavelength $\gg L$) passes through the detector along the $z$ axis, stretching one of the arms by an amount $\delta L(t) \approx h_+(t)L/2$ and compressing the other. In the general case, GWs impinging on the detector will contain both polarizations, so $\delta L(t) \approx [F^+h_+(t) + F^x h_x(t)]L/2$, where $F^+$ and $F^x$ are antenna patterns that weight the contributions from both polarizations according to the position and orientation of the source relative to the detector [14]. These small changes in arm length cause the laser light in the arms to go out of phase with each other. When the beams are recombined, this phase shift shows up in the resulting interference pattern and gives the waveform of the GW signal.

Since the strain $h \ll 1$, it is advantageous to make the detector arms $L$ as large as possible. For example, the LIGO interferometers have arm lengths $L \sim 4$ km. In addition, there are two LIGO sites separated by $\sim 3000$ km to allow for coincidence checks on GW detection, and rule out spurious signals caused by local disturbances such as acoustic noise and laser fluctuations.

## Ground-Based Detectors

Ground-based interferometers with kilometer-scale arms operate in the high frequency range ~ $(1 - 10^4)$ Hz. The lower bound on this frequency range is determined by gravity gradient noise, primarily from seismic activity in the earth. The upper bound arises as the wavelengths of the GWs become comparable to the scale of the detector arms, causing a decrease in sensitivity. The LIGO and Virgo detectors were both planned to be developed in stages: the initial phase would develop full-scale interferometers sensitive enough to detect rare events; the second phase would develop advanced detectors able to make multiple detections per year and be true observational tools. The first generation LIGO and Virgo detectors have reached their design sensitivity and carried out joint science runs. LIGO is currently offline for the upgrade to Advanced LIGO starting in early 2011; early science runs could start ~ 2015 [15]. Upgrades for Advanced Virgo are expected to proceed on a similar timescale. These advanced detectors will be ~ 10 times more sensitive than the initial detectors; for sources with a uniform volume distribution, this implies sensitivity to ~ $10^3$ more sources.

A network of ground-based detectors brings significant improvement in source localization. The Virgo detector in Italy has armlength $L$ ~ 3 km. The Large Cryogenic Gravitational Wave Detector (LCGT) was recently approved for construction in Japan; this will be an underground detector containing two interferometers, each having $L \approx 3$ km. LIGO and Virgo combined give source localization to $\approx 100$ deg$^2$; the addition of the LCGT will reduce this to ~ (5 – 10) deg$^2$. A detector in Australia would improve this to $\approx 1$ deg$^2$ over the entire sky [16].

## Space-Based Detectors

GWs in the source-rich low frequency regime below $f_{GW}$ ~ 1 Hz will be observed with space-borne laser interferometers. The basic idea is to launch a number of spacecraft containing "freely falling" test masses, and use laser interferometry to monitor the changes in light travel time between test masses in each spacecraft caused by passing GWs.

LISA was designed through a collaboration between NASA and ESA, and is the most mature design of the space-borne detectors [17]. LISA comprises three identical spacecraft arranged at the vertices of an equilateral triangle with arm lengths $L$ ~ 5 x $10^6$ km, as shown in Figure 5. Laser signals are sent along the three sides of the triangle to measure changes in distance between test masses inside the spacecraft, due to a passing GW. The entire configuration orbits the sun, following the earth in its orbit by ~ 20 deg.

A key feature of LISA – and, indeed, of any space-based GW detector – is the disturbance reduction system. This is essential for the test masses to be "freely falling" to high enough accuracy that disturbances due to passing GWs are stronger than any other perturbations. Micro-Newton thrusters provide the fine control needed for such drag-free flight. Also, the laser beams spread out over the long distances traveled between spacecraft, resulting in very low photon fluxes at the receiving end.

For this reason, it is not feasible to use standard mirrors for reflecting the light and active mirrors having phase-locked laser transponders will be used instead. See Refs. [8, 17] and references therein for more details.

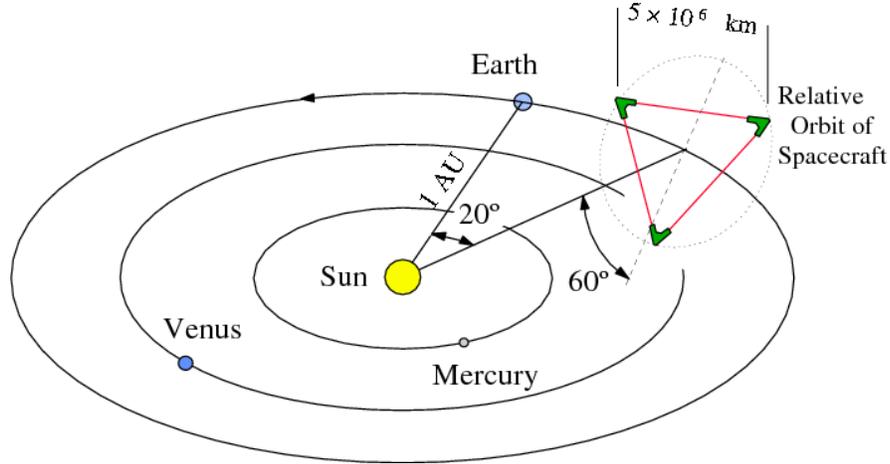

**FIGURE 5.** Schematic drawing of the proposed LISA detector. Three identical spacecraft are located at the vertices of an equilateral triangle to form a laser interferometer with arm length ~ 5 x $10^6$ km. The plane of the triangle is tilted by 60 deg from the ecliptic plane. LISA follows the earth in its orbit about the sun. This figure is from Ref. [18] and is used with permission.

LISA was designed to operate in the frequency band $10^{-4}$ Hz $\leq f_{GW} \leq 10^{-1}$ Hz. Another concept, the proposed Japanese DECi-hertz Interferometer Gravitational-Wave Observatory (DECIGO) would operate at frequencies in the gap between LISA and ground-based detectors, $10^{-1}$ Hz $\leq f_{GW} \leq 10$ Hz. DECIGO also has three drag-free spacecraft, but with significantly shorter arms, $L \sim 10^3$ km. It would be sensitive to the early inspiral stages of double neutron star binaries, as well as intermediate mass black holes, with $M \sim 10^3$ $M_{SUN}$. An even more ambitious future project, the US Big Bang Observer (BBO), would cover the same frequency range as DECIGO, but at greater sensitivity. The BBO design calls for a constellation of three LISA-like configurations, and is aimed at detecting the stochastic GW background from the early universe, and high precision cosmology [8].

## Pulsar Timing Arrays

PTAs are Galactic-sized GW detectors operating in the very low frequency band [19]. A PTA is a collection of millisecond pulsars, with the distance between the earth and each pulsar forming an arm of the detector, as shown in Figure 6. GWs traveling between the earth and a pulsar cause very small changes in the travel time of the pulsar signal. The pulsar signals arriving at earth are received by radio telescopes, and differences in the expected pulse time-of-arrival (TOA) indicate the passage of GWs. PTAs are sensitive to GWs with frequencies in the range $10^{-9}$ Hz $\leq f_{GW} \leq 10^{-6}$ Hz. The lower bound on this range is set by the time span of the observations, typically ~ 10 years. The upper bound arises from the cadence, or time interval between successive observations of each pulsar, typically ~ 1 week.

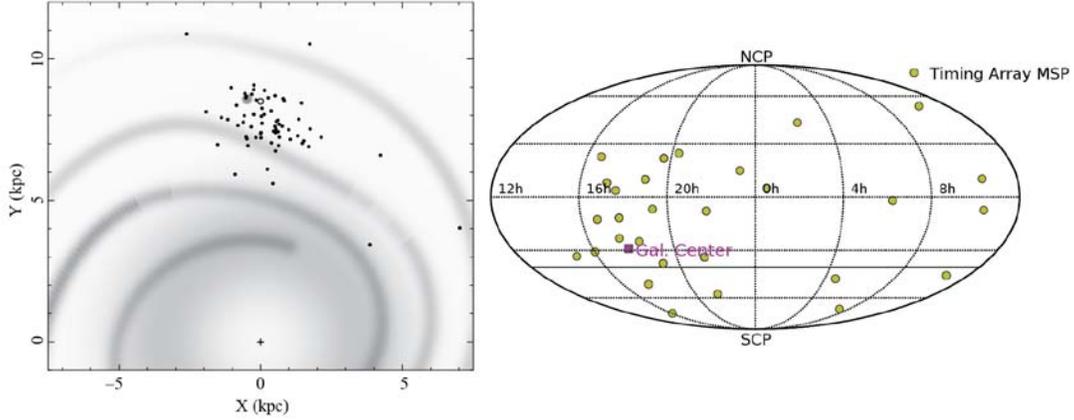

**FIGURE 6.** Millisecond pulsars (msps) used in the International Pulsar Timing Array (IPTA). Left Panel: Black dots give the positions of the IPTA pulsars in the plane of the Galaxy. The + shows the position of the Galactic center, and the open circle the position of the sun. The gray scale shows the free electron density in the plane of the Galaxy, indicating the spiral arms. This figure was produced by R. Manchester, and is used with permission. Right Panel: Distribution of IPTA millisecond pulsars on the sky. The Galactic center is marked with a filled box. This figure was produced by J. Hessels, and is used with permission.

PTAs are possible because millisecond pulsars are extremely stable clocks [20]. These pulsars are believed to be old neutron stars spun up to rotational periods in the range ~ (1 – 10) ms by mass accretion from a companion star. They emit beams of radiation as they rotate. Because their rotational periods are extremely stable, it is possible to time the arrival of these pulses very accurately.

In a PTA, observers use radio telescopes to monitor the root-mean-square (rms) timing residual for each pulsar, which measures the difference between the actual pulsar TOAs and those predicted by a timing model. The timing model accounts for factors such as the pulsar's orbit about a binary companion and dispersion of the pulse by the interstellar medium. GWs passing between a pulsar and the earth cause tiny changes in the TOA, which affects the timing residuals *TR(t)* at time *t* from the initial observation as [9]

$$TR(t) = -\int_0^t H^{ij}(h^e_{ij} - h^p_{ij})\,dt. \qquad (4)$$

Here, $h^e_{ij}$ is the GW strain at the earth when the pulsar is observed; $h^p_{ij}$ is the strain at the pulsar when the EM pulse was emitted (~ $10^3$ years ago, typically); $H_{ij}$ is a geometric term depending on the angle between the earth, pulsar, and GW source; and there is an implied summation over the repeated indices $i,j = 1,2,3$. For more details, see Ref. [21]. From Eq. 3, the GW signal $h$ is very small, and so the expected signal induced by GWs in the TOAs is also tiny, with typical residuals < 100 ns.

GWs passing through the collection of pulsars in a PTA produce correlated changes in the TOAs from different pulsars. An ideal PTA would have millisecond pulsars with sufficiently small timing residuals evenly distributed on the sky. Overall, one needs to time at least 20 pulsars, each having a rms timing residual $\lesssim$ 100 ns, for at least ~ 5 years to detect a stochastic GW background in the nHz regime; the number of pulsars, needed precisions, and time to detection could be larger [22]. Detection of GWs from individual SMBH binaries will require rms timing residuals < 10 ns. See Refs. [23, 24].

Currently there are three pulsar timing projects: the Parkes Pulsar Timing Array (PPTA) in Australia; the North American Nanohertz Observatory for Gravitational Waves (NANOGrav); and the European Pulsar Timing Array (EPTA). Overall, at least 37 millisecond pulsars are being timed for use in PTAs; most of these are being monitored by more than one PTA group, and about half of them currently have timing residuals < 500 ns. The individual PTA groups are working on arrangements to combine their datasets and form the International Pulsar Timing Array (IPTA) [9].

## GRAVITATIONAL WAVE SOURCES

This section presents an overview of astrophysical sources of GWs. We focus mainly on various types of binaries, which are expected to be the strongest and most likely sources of GWs in the high, low, and very low frequency regimes. For more details and information on other sources, see the excellent reviews in Refs. [6] and [12].

### General Properties of Binary GW Sources

A mass-radius diagram, shown in Figure 7, provides a convenient way to understand key properties of GW sources [6]. Here, the vertical axis gives the scale of the radiating system, and the horizontal axis its total mass. The lines and symbols on the plot show order-of-magnitude estimates and relationships.

The three parallel dark blue lines denote lines of constant frequency $f_{GW}$, and delimit the high (earth band) and low (space band) frequency regions of the plot; sources lying within these regions can be detected by ground-based and space-based interferometers, respectively. The steeper solid gray line gives the black hole radius, $R_{Schw} = 2GM/c^2$; no systems can exist below this line, as then they would be smaller than their Schwarzschild radii. Note that from the ground, only GWs from ~ stellar mass objects can be observed, whereas GWs from massive black holes can be detected from space.

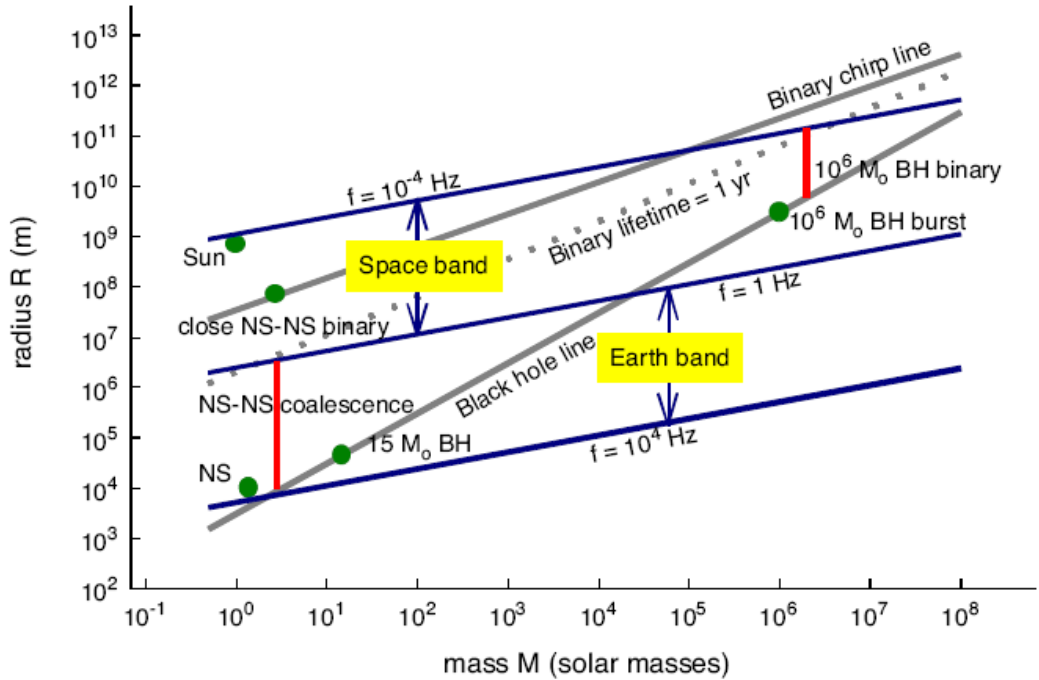

**FIGURE 7.** Mass-radius diagram for GW sources. The low frequency (space band) and high frequency (earth band) regions are shown, along with relationships and constraints on the sources. This figure is from Ref. [6] and is used with permission.

Recall that GWs carry energy and momentum away from the binary. This causes the binary orbit to shrink, as the orbital and GW frequencies increase with time. The timescale on which the binary orbits change under GW emission $\sim a^4$, where $a$ is the separation between the binary components; this means that the orbits shrink faster as the separation decreases.

Two additional diagonal lines in Figure 7 mark relationships pertaining to binary evolution; here, $R$ gives the binary separation $a$. The dashed gray line shows binaries that are $\sim$ one year from final coalescence. Binaries below this line will merge during one year of observation, generally producing a final burst of gravitational radiation. The solid gray line labeled "binary chirp line" denotes the region where the shrinking (chirping) of the binary orbit can be detected during a year. That is, binaries above this line will not appear to exhibit any frequency evolution during roughly a year of observation.

Look now at some specific GW sources on this plot. First, the vertical line in Figure 7 labeled "NS-NS coalescence" represents a binary that is within roughly the last several thousand orbits (and GW cycles) before final coalescence. This system evolves at nearly constant total mass $M$ (the actual mass change due to gravitational radiation is much too small to be detectable on this plot) to smaller separations and higher frequencies, as shown by the vertical red line. Such NS coalescences are key sources for ground-based detectors. Since it is below the dotted line, this system will evolve in < 1 year; in fact, the last few thousand cycles that will be observed by ground-based detectors last ~ 3 minutes. Next, the filled circle labeled "close NS-NS binary" is at a wider separation and an earlier stage in its evolution; as plotted here, it is right on the binary chirp line, so the frequency evolution of this source is just barely detectable in one year of observation by a space-based detector. Finally, the red vertical line at the upper right part of the diagram shows the frequency evolution of a MBH binary with total mass $M \sim 10^6$ $M_{SUN}$. This system enters the low frequency band ~ 1 year before its final coalescence, and evolves upwards in frequency until the individual black holes merge to form a single final black hole. This evolution will be easily detectable by a space-based interferometer such as LISA.

## Overview of GW Source Detections and Rates

How many GW sources will be detected? The answer to this question depends on the type of source and detector. In addition, estimates of the detection rates are generally based on extrapolations of model calculations with varying uncertainties. In this section we present an overview of current expectations for detections of GWs from binaries, focusing on values considered realistic and robust by the GW astrophysics community.

First generation ground-based detectors have already carried out science data-taking runs in the high frequency regime. In order for these initial instruments to make a detection, they would need to operate for ~ 5 years (in the most optimistic case) to 20 years or more (in more realistic scenarios for compact binary formation), due to the limited volume of space within which they can receive detectable signals [25]. The second-generation, or advanced, detectors are expected to make the first firm detections of GWs from compact objects within the next ~ 5 years. Specifically, Advanced LIGO is predicted to detect between ~ (0.4 – 400) NS binaries per year, with the most realistic estimate being ~ 40 per year. For stellar black hole binaries (with each black hole having mass ~ $10 M_{SUN}$), the rate spans the region ~ (0.2 – 300) detections per year, most realistically ~ 20 per year. And for NS-black hole binaries, the rates range from ~ (0.2 – 300) per year, with the most realistic estimate being ~ 10 per year. Advanced LIGO may also detect stellar black holes with mass ~ 10 $M_{SUN}$ spiraling into a larger black hole with mass ~ 100 $M_{SUN}$ at the rate of ~ 10/year. All these rates are from Ref. [25].

Since the low frequency regime is much richer in astrophysical sources, we can expect many more detections. The actual rates depend on the specific design of the space-based interferometer; for our purposes here, we consider the baseline LISA design and science, as described in Refs. [6, 26, 27]. LISA would detect MBH binary mergers with masses $M \sim (10^4 - 10^7) M_{SUN}$ at rates of ~ 1 per year at redshift $z < 2$ and ~ 30 per year out to $z \sim 15$. These sources would be observed with relatively high amplitude signal-to-noise, allowing high precision extraction of the source parameters from the waveforms, measuring masses and spins to ~ (1 – 2)% or better, and luminosity distances to ~ a few percent. LISA would also detect stellar black holes of mass ~ 10 $M_{SUN}$ spiraling into a MBH with $M \sim (10^4 - 10^7) M_{SUN}$ at the rate of ~ 50 per year; observations of these sources would allow precision tests of general relativity through mapping of the Kerr spacetime metric, and would determine the masses and spins of the black holes to ~ 0.1% accuracy. Finally, LISA would detect GWs from ~ 20,000 individual compact binaries, mostly WD binaries in the Galaxy; many more compact binaries form a confusion-limited foreground source.

The total population of MBH binaries and SMBH binaries produces a stochastic background of GWs extending over a wide range of frequencies. In the very low frequency regime, the main contributors to this background are SMBH binaries, with masses $M \geq 10^8 M_{SUN}$, at low redshifts $z \leq 2$. This is a key source for PTAs, and could be detected in the next ~ (5 – 10) years by using combined datasets in the IPTA [9]. This will require improving the timing through various means, including discovering new millisecond pulsars, better correcting for interstellar medium effects or observing at higher frequencies, and increasing the bandwidth and/or observing time. The advent of new radio telescopes, including the Square Kilometer Array (SKA) and its precursors, will bring new levels of precision to very low frequency GW observations, and allow detection of individual sources and characterization of various stochastic backgrounds [9, 28].

## THE FUTURE: MULTI-MESSENGER ASTROPHYSICS

Throughout the history of astronomy, opening a new window onto the universe has always been accompanied by scientific progress, discovery, and excitement. The new information made available is often synergistic with and complementary to data from existing techniques, allowing progress in scientific areas already under investigation. Perhaps the greatest opportunities arise in discovery, from unexpected features of sources already being studied by other means, to completely new phenomena. The intellectual excitement surrounding all these developments invigorates both the science and the scientists, with bursts of creativity and new ways of thinking.

So, we can look forward with great anticipation as the GW window onto the universe opens this decade [29, 30]. Given the current status of GW detectors, this opening will come in stages. The high frequency portion of the GW spectrum will likely open first, with the first detections of NS binaries and stellar black hole binaries by the advanced ground-based interferometers expected to start ~ 2015. The very low frequency regime will likely come next, as PTAs reach the required sensitivities for detection of GW backgrounds in ~ 5 – 10 years. Opening the low frequency region with a space-borne detector will follow in the next decade, with detections of MBH binaries leading the way.

These developments are particularly exciting since they are occurring in the context of new instruments and directions developing in EM astronomy [31]. Wide FOV instruments and surveys are especially well-suited to looking for EM counterparts to detections of GWs from merging binaries ranging from NS to MBH binary coalescences. GWs from MBH binaries at high redshifts will bring key information on the merger history of MBHs (and, by extension, their host galaxies) through cosmic time, especially during the era of reionization [32]. And searches for GWs from transient events discovered in EM surveys can potentially unlock the mechanisms of the engines powering these sources [33].

In fact, combined EM-GW studies have already begun, using data from the initial ground-based GW detectors. Figure 8 shows M31 (the Andromeda Galaxy) with the overlaid error box localizing the position of the short duration, hard spectrum GRB 070201 [34]. The LIGO-Hanford detectors H1 (4 km arms) and H2 (another interferometer in the same vacuum system, having 2 km arms) were operating and taking science-quality data during this event. Analysis of this data found no plausible GW candidates within a 180 s long window around the time of GRB 070201. This rules out a compact binary, having component masses in the range $1\ M_{SUN} < m_1 < 3\ M_{SUN}$ and $1\ M_{SUN} < m_2 < 40\ M_{SUN}$ and location in the spiral arms of M31, as the progenitor of GRB 070201 at $> 99\%$ confidence. In addition, if GRB 070201 were caused by a binary NS merger, this analysis excludes a source at distance $r < 3.5$ Mpc, assuming random inclination, at 90% confidence.

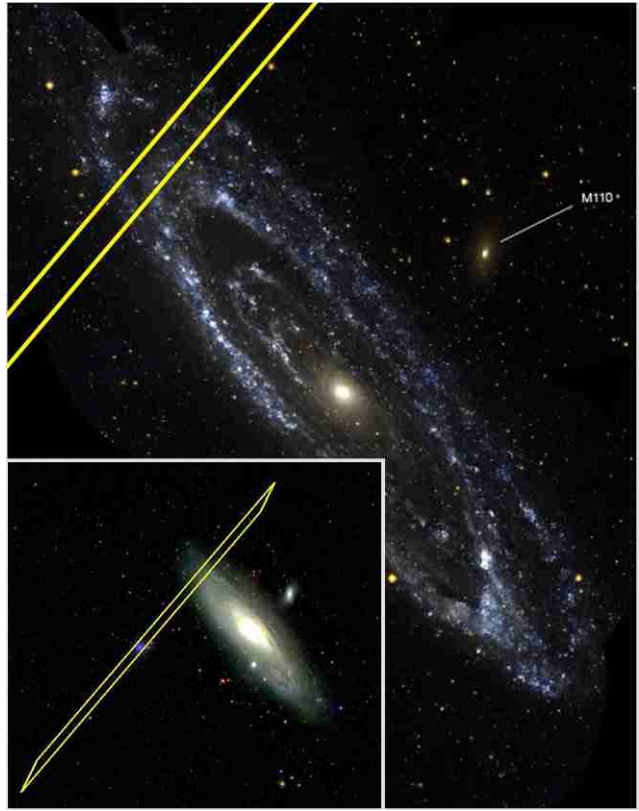

**FIGURE 8.** M31 with overlapping error box for the position of the short duration, hard spectrum GRB 070201. The main figure shows the overlap of the error box and the spiral arms of M31 in UV light [35]; the inset shows the full error box overlaid on an SDSS image of M31 [36, 37]. Figure reprinted from Ref. [34], reproduced by permission of the AAS.

Figure 8 gives a tantalizing hint of the scientific possibilities that will arise as the GW window onto the universe opens wide. GW astronomy will grow beyond the relatively small current GW community, becoming a vital tool for understanding the universe and involving the much larger astronomy and astrophysics community. Exciting times lie ahead, indeed!

## ACKNOWLEDGMENTS

Special thanks go to P. Tyler for producing Figure 3, and R. Manchester and J. Hessels for producing the panels in Figure 6. It is a pleasure to acknowledge helpful comments from my colleagues M. McLaughlin, J. Livas, and I. Thorpe. This work was supported in part by NASA grant 09-ATP09-0136.